\documentclass[fleqn,usenatbib,useAMS]{mnras}
\usepackage{graphicx}
\usepackage{bm,amsmath,amssymb, ulem}
\newcommand{\pd}[2]{\frac{\partial {#1}}{\partial {#2}}}
\newcommand{\od}[2]{\frac{\mathrm{d} {#1}}{\mathrm{d} {#2}}}

\title[CR heating in the early universe]{Resistive Heating Induced by Streaming Cosmic Rays Around a Galaxy in the Early Universe}
\author[S. L. Yokoyama and Y. Ohira]{Shota L. Yokoyama $^{1}$
    \thanks{E-mail: \href{mailto:s_yokoyama@eps.s.u-tokyo.ac.jp}{s\_yokoyama@eps.s.u-tokyo.ac.jp}},
    Yutaka Ohira $^{1}$
    \\
    $^{1}$ Department of Earth and Planetary Science, The University of Tokyo, 7-3-1 Hongo, Bunkyo-ku, Tokyo 113-0033, Japan
}

\date{Accepted XXX. Received YYY; in original form ZZZ}
\pubyear{2023}

\begin{document}
\label{firstpage}
\pagerange{\pageref{firstpage}--\pageref{lastpage}}
\maketitle

\begin{abstract}
    It is expected that cosmic rays (CRs) escape from high-redshift galaxies at redshift $z\sim 10 \, - \, 20$ because CRs are accelerated by supernova remnants of the first stars.
    Although ultraviolet and X-ray photons are widely considered the main source of heating of the intergalactic medium, CRs can also contribute to it.
    When the CRs propagate in the intergalactic medium, in addition to the heating process due to CR ionization, resistive heating occurs due to the electron return current induced by the streaming CRs.
    We evaluate the heating rate around a galaxy as a function of the distance from the galaxy.
    We find that the resistive heating induced by CRs dominates over the other heating processes in the vicinity of the galaxy $r \lesssim 10^2 \, \mathrm{kpc}$ until the temperature reaches $T\sim 10^4 \, \mathrm{K}$.
    We also recalculate the strength of the magnetic field generated by streaming CRs under the presence of X-ray heating and show that achieved strength can be about $1$ order of magnitude smaller when the X-ray heating is included.
    The presence of the "first" CRs could be confirmed from the characteristic signature of CR heating imprinted on the $21$-$\mathrm{cm}$ line map in future radio observations.
\end{abstract}

\begin{keywords}
cosmic rays -- magnetic fields -- plasmas -- intergalactic medium
\end{keywords}

\section{Introduction}
\label{sec:1}
Our Universe experienced the transition from a cold and neutral state to a hot and ionized state sometime in cosmic history.
This transition phase is called the epoch of reionization (EoR) and is considered to be caused by the radiation from stars and galaxies.
Because the first stars and galaxies are formed at redshift $z\sim 10 - 30$ \citep[e.g.][]{Bromm2004, Bromm2011}, the EoR begins around this era.
The thermal and reionization history of the intergalactic medium (IGM) can be touched with some observations \citep[see review for][]{McQuinn2016}.
Lyman $\alpha$ forest observations indicate that the hydrogen reionization ended at $z \sim 6$.
Thomson scattering optical depth for the cosmic microwave background (CMB) photons also gives the measure for the reionization, and it suggests that the Universe became ionized at $z\sim 10$ \citep[][]{Planck2020}.

In the standard picture, the reionization is accomplished by the ultraviolet (UV) photons from high-$z$ galaxies, while X-rays also contribute to the ionization.
X-rays drive the global heating of the IGM because they can propagate longer distances than UV photons.
However, the nature of X-ray emission in the early Universe is still highly unknown \citep[][]{McQuinn2012}.

In addition to photo-heating, heating by cosmic rays (CRs) is promising to raise the IGM temperature.
The standard scenario of CR acceleration in the current Universe is that a shock wave accelerates CRs in supernova remnants \citep[e.g.][]{Blasi2013}.
In the same way in the current galaxies, it is implied that supernova remnants of the first stars also accelerate CRs in spite of the weakness of the magnetic field responsible for CR acceleration \citep[][]{Ohira2019}.
Because the first supernova remnants emerge at $z \sim 20$, CRs accelerated in them may take part in the heating of the IGM.

IGM heating by CRs is discussed in a few papers.
Direct heating by CRs, which is caused by Coulomb interaction between CRs and residual free electrons and by ionization of neutral atoms, is discussed in \citet{Sazonov2015, Leite2017} and global temperature increase of $\Delta T=10 \sim 200 \, \mathrm{K}$ is inferred.
\citet{Miniati2011} discussed the resistive heating induced by streaming CRs while their main concern was magnetic field generation.
Although they calculated the temperature increase by resistive heating around a galaxy, they did not include the other heating mechanisms.
Our previous work showed that the heating rate by resistive heating can be significantly higher than that by direct heating \citep[][]{Yokoyama2022}.
In this work, we reconsider the role of CRs in the local heating of IGM around an individual galaxy, including both direct and resistive heating, and compare them with the other heating processes.
It will be shown that CR resistive heating can be dominant in the vicinity of the galaxy until the temperature reaches $T\sim 10^4 \, \mathrm{K}$, while direct heating contributes to the heating far outside the galaxy.
In addition, we recalculate the magnetic fields generated by the mechanism of \citet{Miniati2011} under the presence of X-ray heating and demonstrate that the field strength becomes about $1$ order of magnitude weaker than that in the absence of X-ray heating.

Recently, James Webb Space Telescope (JWST) begins to shed light on the nature of high-$z$ galaxies \citep[e.g.][]{Robertson2022}.
Using the early data of JWST, the possibility of the higher star formation efficiency and excess of bright galaxies in a high-$z$ environment has been already pointed out \citep[e.g.][]{Harikane2022}.
Because the production of CRs in galaxies is certainly correlated with star formation, this implies high CR production efficiency and motivates us to understand the impact of CRs on the evolution of the early Universe.
The heating of the IGM can be investigated by radio observations of $21$-$\mathrm{cm}$ hydrogen line.
$21$-$\mathrm{cm}$ line reflects the gas temperature through the spin temperature.
The difference between the spin and CMB temperature is observed as the emission or absorption against CMB radiation \citep[e.g.][]{Furlanetto2006}.
The Experiment to Detect the Global EoR Signature (EDGES) claimed that the absorption much deeper than the prediction of theoretical modeling was detected \citep[][]{Bowman2018}, while the SARAS 3 experiment reported the non-detection of it \citep[][]{Bevins2022}.
Future radio observations such as Square Kilometer Array (SKA) will solve this conflict and reveal the properties of IGM in different redshifts \citep[e.g.][]{Koopmans2015}.

The rest of this paper is organized as follows.
In Section \ref{sec:2}, we explain how streaming CRs induce resistive heating.
Various sources and processes of IGM heating are discussed in Section \ref{sec:3}, where a comparison between the resistive heating and the others is made.
Discussion is given in Section \ref{sec:4}, followed by the conclusion in Section \ref{sec:5}.


\section{Physics}
\label{sec:2}
Before discussing the heating around a galaxy, it is instructive to review the physical mechanism of resistive heating.
We consider the system where high-energy CRs stream through a thermal medium, that is, three-component plasma which consists of thermal protons (subscript $\mathrm{p}$), thermal electrons (subscript $\mathrm{e}$), and nonthermal CR protons (subscript $\mathrm{CR}$).
In order to cancel the electric current $\bm{J}_{\rm CR}$ carried by the streaming CRs, a flow of thermal electrons is induced, that is, the electric current
$\bm{J}_{\rm th}=-\bm{J}_{\rm CR}$ is produced.
Although the friction between CRs and thermal particles is negligible, the collision between thermal electrons and thermal protons is not negligible and induces a resistive electric field.
When the collision between thermal components is mediated by Coulomb collision, the resistivity is given by the so-called Spitzer resistivity:
\begin{equation}
    \eta_{\rm C} = 7.23 \times 10^{-9} \log \Lambda \left( \frac{T}{1 \, \mathrm{K}} \right)^{-3/2} \, \mathrm{s},
    \label{eq:resistivity}
\end{equation}
where $T$ is the temperature and $\log \Lambda$ is called the Coulomb logarithm, and we assume $\log \Lambda = 20$ hereafter.
This choice is not sensitive to the following discussion because of its logarithmic dependence.
Then, the induced electric field is given by Ohm's law
\begin{equation}
    \bm{E} = \eta_{\rm C} \bm{J}_{\rm th} = - \eta_{\rm C} \bm{J}_{\rm CR}.
    \label{eq:ohm}
\end{equation}
This electric field induces the Joule heating of the plasma.
For the cosmological time scale considered in this work, soon after the electrons are heated by the Joule heating, they exchange their energy to protons and neutral hydrogens, and thermal equilibrium between them is quickly achieved.
Then, the temperature evolution of the gas by this heating mechanism is described by the following energy equation:
\begin{equation}
    \od{}{t} \left( \frac{3}{2} (1 + \chi_{\rm e}) n_{\rm H} k_{\rm B} T \right) = \eta_{\rm C} J_{\rm CR}^2.
    \label{eq:temperature}
\end{equation}
$n_{\rm H}$ is the number density of hydrogen including ionized and neutral ones.
In this paper, we consider the gas composed only of hydrogen, for simplicity.
We also introduce the electron fraction $\chi_{\rm e}$ and describe the electron and neutral hydrogen number density as $n_{\rm e} = \chi_{\rm e} n_{\rm H}$ and $n_{\rm H_{\rm I}} = (1-\chi_{\rm e}) n_{\rm H}$, respectively.
Equation (\ref{eq:temperature}) describes the resistive heating induced by streaming CRs and the heating rate depends on the square of CR current density.
In the next section, we estimate the heating rate for the parameters around a galaxy and compare it with the other heating mechanisms.

\section{Heating around a galaxy}
\label{sec:3}
We here discuss the local heating around a
galaxy, 
simply assuming that the propagation of CRs is free streaming.
This work will give a rough estimate of the heating rate and temperature increase around a galaxy.

In this section, we first discuss the sources of photons and CRs and estimate their luminosities in \ref{subsec31}.
Then, we enumerate the possible heating mechanisms and the equations describing the heating rates in \ref{subsec32}.
The estimated heating rates and temperature increase are given in \ref{subsec33} as a function of distance from a galaxy.
Magnetic field generation is discussed in \ref{subsec34}.

\subsection{Photons and CRs from a galaxy}
\label{subsec31}

\subsubsection{UV photons}
The stars formed in the high-$z$ galaxy emit photons with UV energy.
These photons ionize and heat the surrounding medium, making the H$_{\, {\rm II}}$ region around a galaxy.
The maximum radius of this H$_{\, {\rm II}}$ region is roughly estimated by neglecting the recombination and by balancing the UV photon number and hydrogen number included in the H$_{\, {\rm II}}$ region.
This makes the following estimate:
\begin{equation}
    r_{{\rm H}_{\, {\rm II}}, \rm{max}} = \left( \frac{3 N_{\gamma}}{4 \pi n_{{\rm H}}} \right)^{1/3} \sim 10 \, \mathrm{kpc},
    \label{eq:HII}
\end{equation}
where the radius is evaluated at $z\sim 10$ and measured in the proper frame \citep[e.g.][]{Cen2006, Loeb2010}.
$N_{\gamma}$ is the total number of ionizing photons that escaped from the galaxy and it is estimated in \cite{Loeb2010}.
Because we neglected the recombination, the actual radius does not reach this value.
Hereafter, we consider the heating of a cold and almost neutral region outside the H$_{\, {\rm II}}$ region $r > r_{{\rm H}_{\, {\rm II}}} \sim 1 \, \mathrm{kpc}$.
We do not consider UV heating, assuming that UV photons cannot penetrate into the region outside $r_{\rm H_{II}}$ and do not contribute to heating.

\subsubsection{X-ray photons}
Although X-rays are widely recognized as the main contributor to the global heating of IGM, there remains large uncertainty about the X-ray emitters in the early Universe.
\citet{Furlanetto2006} parameterized the luminosity of X-ray $L_{\rm X}$ as a function of star formation rate ($\mathrm{SFR}$), based on the correlation between them in the nearby starburst galaxies:
\begin{equation}
    L_{\rm X} = 3.4 \times 10^{40} \, \mathrm{erg \, s^{-1}} \left( \frac{f_{\rm X}}{1} \right) \left( \frac{\mathrm{SFR}}{1 \, \mathrm{M_{\odot}} \, \mathrm{yr}^{-1}} \right),
    \label{eq:Xluminosity}
\end{equation}
Here, $f_{\rm X}$ is introduced as a correction factor that covers the large uncertainty about the nature of the X-ray sources in the early Universe.
There are several candidates for X-ray sources.
In the current Universe, high-mass X-ray binaries produce the bulk of high-energy photons of nearby galaxies.
Although the number of X-ray binaries in the high-$z$ galaxies depends on the initial mass functions and metallicity, they can also be a dominant source of galactic X-rays \citep[][]{Furlanetto2006, Mirabel2011}.
Inverse Compton scattering of CMB by nonthermal electrons which are accelerated in the supernova remnants is another source of X-rays.
Thermal emission from the supernova remnant shock can also produce soft X-ray photons \citep[][]{Johnson2011, McQuinn2012}.
In addition, active galactic nuclei may contribute to the X-ray emission from galaxies \citep[][]{McQuinn2012}.
An even more exotic contribution is expected if we consider the annihilation of dark matter \citep[][]{Belikov2009}.
Because of this complexity, we adopt the simplified relation introduced by (\ref{eq:Xluminosity}) and impose all the uncertainties to the factor $f_{\rm X}$.

Because the cross-section for photoionization depends on the frequency $\nu$ of the incident X-ray, we assume a power-law spectrum $L_{\nu} \propto \nu^{-\alpha}$ with $\alpha = 1.5$ \citep[][]{Rephaeli1995, Furlanetto2006}.
The spectrum is normalized so that the integral over $\nu$ gives the total X-ray luminosity $L_{\rm X}$ given in Equation (\ref{eq:Xluminosity}).
%
%
We introduce the minimum and maximum X-ray energy as $h\nu_{\rm min}=0.2 \, \mathrm{keV}$, and $h\nu_{\rm max} = 10 \, \mathrm{keV}$, respectively, assuming that soft X-rays with $h\nu \lesssim 0.2 \, \mathrm{keV}$ are consumed in forming the H$_{\, {\rm II}}$ region and do not propagate to $r > 1 \, \mathrm{kpc}$.

\subsubsection{Cosmic rays}
\label{subsubsec313}
The most favored mechanism to accelerate CRs is diffusive shock acceleration (DSA) \citep[][]{Axford1977, Krymsky1977, Bell1978, Blandford1978}.
In order for DSA to operate, it requires the existence of a collisionless shock wave and magnetic turbulence which scatters the CRs.
Because the strength of the magnetic field is highly unknown, whether DSA works in the early Universe is not trivial.
\citet{Ohira2019} investigated the supernova remnants (SNRs) of the first stars and accretion shocks, which are candidates of acceleration systems in the early Universe, and showed that the SNR shock is promising to accelerate the first CRs.
They predicted that the first CRs 
are accelerated to energies in the range of
$3 \,  \mathrm{MeV} \lesssim E \lesssim 3 \, \mathrm{GeV}$.
Assuming that $10$ per cent of the kinetic energy of the SNR is converted to that of CRs, corresponding to conversion efficiency $\varepsilon_{\rm CR}=0.1$, we can link the CR luminosity with star formation rate:
\begin{align}
    L_{\rm CR} = 3.2 \times 10^{40} & \, \mathrm{erg \, s^{-1}} \left( \frac{\varepsilon_{\rm CR}}{0.1} \right) \nonumber \\
    & \left( \frac{\mathrm{SFR}}{1 \, \mathrm{M_{\odot}} \, \mathrm{yr}^{-1}} \right) \left( \frac{E_{\rm SN} \nu_{\rm SN}}{10^{49} \mathrm{erg} \, \mathrm{M_{\odot}}^{-1}} \right),
    \label{eq:CRluminosity}
\end{align}
Here we assumed that each supernova liberates energy of $E_{\rm SN} = 10^{51} \, \mathrm{erg}$ 
at a rate $\nu_{\rm SN} = 0.01 \, \mathrm{M_{\odot}}^{-1}$, although all the parameters $\varepsilon_{\rm CR}$, $E_{\rm SN}$, and $\nu_{\rm SN}$ are also uncertain in the early Universe.
The average energy of supernova $E_{\rm SN}$ and supernova rate $\nu_{\rm SN}$ might be higher if the initial mass function in the first galaxy is a top-heavy distribution with POP-III stars.
In addition, recent observations by JWST imply that the star formation rate is higher than $1 \, \mathrm{M_{\rm \odot} yr^{-1}}$ in the high-$z$ galaxies \citep[e.g.][]{Naidu2022}.

In the same way as X-rays, we consider the power-law momentum distribution of CRs accelerated in SNRs:
\begin{equation}
    Q(p) = \frac{\mathrm{d}^2 N}{\mathrm{d} p \, \mathrm{d} t} = Q_0 p^{-s}, \quad p = \frac{\gamma v}{c}.
    \label{eq:CRspectrum}
\end{equation}
$Q(p)$ gives the number of CRs supplied by the galaxy in unit time in unit momentum.
We introduced normalized momentum $p=(\gamma m_{\rm p} v)/(m_{\rm p} c)$, where $v$ is the velocity and $\gamma$ is the corresponding Lorentz factor.
Once we further define normalized velocity $\beta = v/c$, $\beta$ and $\gamma$ are written in terms of normalized momentum $p$, that is, $\beta = p/\sqrt{1+p^2}$ and $\gamma = \sqrt{1+p^2}$.
In Equation (\ref{eq:CRspectrum}), $s$ is the spectral index and DSA predicts $s=2$ \citep[e.g.][]{Drury1983}, while the observations of CRs at the Earth suggest $s \simeq 2.4$.
The spectral index is modified to a slightly larger value than $s=2$ by the escape process from the acceleration region \citep[][]{Ohira2010a}.
Then, the  spectral index of escaped CRs is expected to be $s = 2 - 2.4$.
We use the minimum and maximum momentum $p_{\rm min}, \, p_{\rm max}$ corresponding to the energy $E_{\rm min} = 3 \, \mathrm{MeV}$ and $E_{\rm max} = 3 \, \mathrm{GeV}$, respectively, as fiducial values \citep[][]{Ohira2019}.
%
%

The momentum distribution at the radius $r$ is given by the steady-state solution of the equation of continuity with the source term $Q(p)$.
The normalization factor $Q_0$ is determined by requiring that the total luminosity corresponds to that given by Equation (\ref{eq:CRluminosity}).
Therefore,
\begin{equation}
    \frac{\mathrm{d} n}{\mathrm{d} p} = \frac{Q_0}{4\pi r^2 v} p^{-s}, \quad
    L_{\rm CR} = \int_{p_{\rm min}}^{p_{\rm max}} (\gamma-1) m_{\rm p} c^2 Q_0 p^{-s} \mathrm{d} p.
\end{equation}

\subsection{Heating processes}
\label{subsec32}
In this subsection, we enumerate various heating mechanisms and derive equations to evaluate the heating rate of each process.
Here we assume that X-rays and CRs are emitted from a point source galaxy with the luminosity estimated by (\ref{eq:Xluminosity}) and (\ref{eq:CRluminosity}).
The energy flux of them declines as a function of the distance $r$ from the galaxy.

\subsubsection{X-ray heating}
When the X-rays penetrate into the IGM, neutral hydrogen is photoionized.
The cross-section for photoionization is given by
\begin{equation}
    \sigma_{\nu} = \sigma_0 (h\nu / I_{\rm H})^{-3},
    \label{eq:cross-section}
\end{equation}
where $\sigma_0 = 6.3 \times 10^{-18} \, \mathrm{cm^{-3}}$ and $I_{\rm H} = 13.6 \, \mathrm{eV}$ is the ionization potential of hydrogen \citep[e.g.][]{Draine2011}.
The secondary electrons emitted by photoionization can heat and further ionize the gas.
About $20$ per cent ($\eta_{\rm X} = 0.2$) of electron energy is converted into the heat \citep[][]{Furlanetto2010}.
Because the energy dependence of $\eta_{\rm X}$ changes the value only by a factor of $\sim 2$, here we assume it is constant against $\nu$ \citep[][]{Furlanetto2010}.
Then, the heating rate density (unit of $\mathrm{erg \, cm^{-3} \, s^{-1}}$) by X-ray is obtained as follows:
\begin{align}
     \varepsilon_{\rm X} = n_{\rm H_{\rm I}} \int_{\nu_{\rm min}}^{\nu_{\rm max}} \frac{\eta_{\rm X} L_{\nu}}{4\pi r^2 h \nu} & \sigma_{\nu} (h\nu - I_{\rm H}) e^{-\tau_{\nu}} \mathrm{d} \nu.
    \label{eq:Xray}
\end{align}
$r^{-2}$ factor describes the decrease of X-ray photons as going away from the galaxy.
We included the absorption effect by $e^{-\tau_{\nu}}$, where $\tau_{\nu}=n_{\rm H_{\rm I}} \sigma_{\nu} r$ is the optical depth.
%
%

\subsubsection{CR resistive heating}
As explained in Section \ref{sec:2}, electron return current induced by streaming CRs heat the gas resistively.
The heating rate density by CR resistive heating is given by Equation (\ref{eq:temperature})
\begin{equation}
    \varepsilon_{\rm CR, res} = \eta_{\rm C} J_{\rm CR}^2.
    \label{eq:CRresistive}
\end{equation}
and the CR current density is given by
\begin{equation}
    J_{\rm CR} = ec \int_{p_{\rm min}}^{p_{\rm max}} \beta \od{n}{p} \mathrm{d} p.
    \label{eq:CRcurrent}
\end{equation}
%
%
%

\subsubsection{CR direct heating}
In addition to resistive heating, CRs can directly interact with free electrons by Coulomb interaction and with neutral hydrogen by ionization.
Both interactions result in the heating of gas and the heating rates of a CR proton are given for Coulomb heating by
\begin{equation}
    \left( \od{E}{t} \right)_{\rm C} = \frac{4\pi e^4 n_{\rm e}}{m_{\rm e} \beta c} \left[ \ln \left( \frac{2m_{\rm e} c^2 \beta}{\hbar \omega_{\rm pe}} \right) - \frac{\beta^2}{2} \right],
    \label{eq:lossCoulomb}
\end{equation}
and for ionization heating by
\begin{equation}
    \left( \od{E}{t} \right)_{\rm I} = \frac{4\pi e^4 n_{\rm H_{\rm I}}}{m_{\rm e} \beta c} \left[ \ln \left( \frac{2m_{\rm e} c^2}{I_{\rm H}} p^2  \right) - \beta^2 \right],
    \label{eq:lossIon}
\end{equation}
respectively \citep[][]{Enblin2007, Leite2017}.
$\hbar$ is the reduced Planck constant and $\omega_{\rm pe} = (4\pi n_e e^2/m_e)^{1/2}$ is the electron plasma frequency.
We can calculate the total heating rate density of CR direct heating by integrating the sum of (\ref{eq:lossCoulomb}) and (\ref{eq:lossIon}) weighted with the CR distribution:
\begin{equation}
    \varepsilon_{\rm CR, dir} = \int_{p_{\rm min}}^{p_{\rm max}} \left[ \left( \od{E}{t} \right)_{\rm C} + \left( \od{E}{t} \right)_{\rm I} \right] \od{n}{p} \mathrm{d} p.
    \label{eq:CRdirect}
\end{equation}

\subsubsection{Ohter heating mechanisms}
Although we omit the detailed discussion, we mention the other possibilities for heating processes.
IGM heating by structure formation shock has the potential to heat the gas significantly \citep[][]{Furlanetto2004, Jia2020}.
Plasma instabilities caused by $\mathrm{TeV}$ blazers may also contribute to the IGM heating \citep[][]{Schlickeiser2012, Chang2012}.
Because these heating processes are fairly unknown and difficult to discuss within our model, we postpone the discussion about them to future works.

\subsection{Heating rate and temperature evolution}
\label{subsec33}
\begin{figure}
    \centering
    \includegraphics[width=80mm]{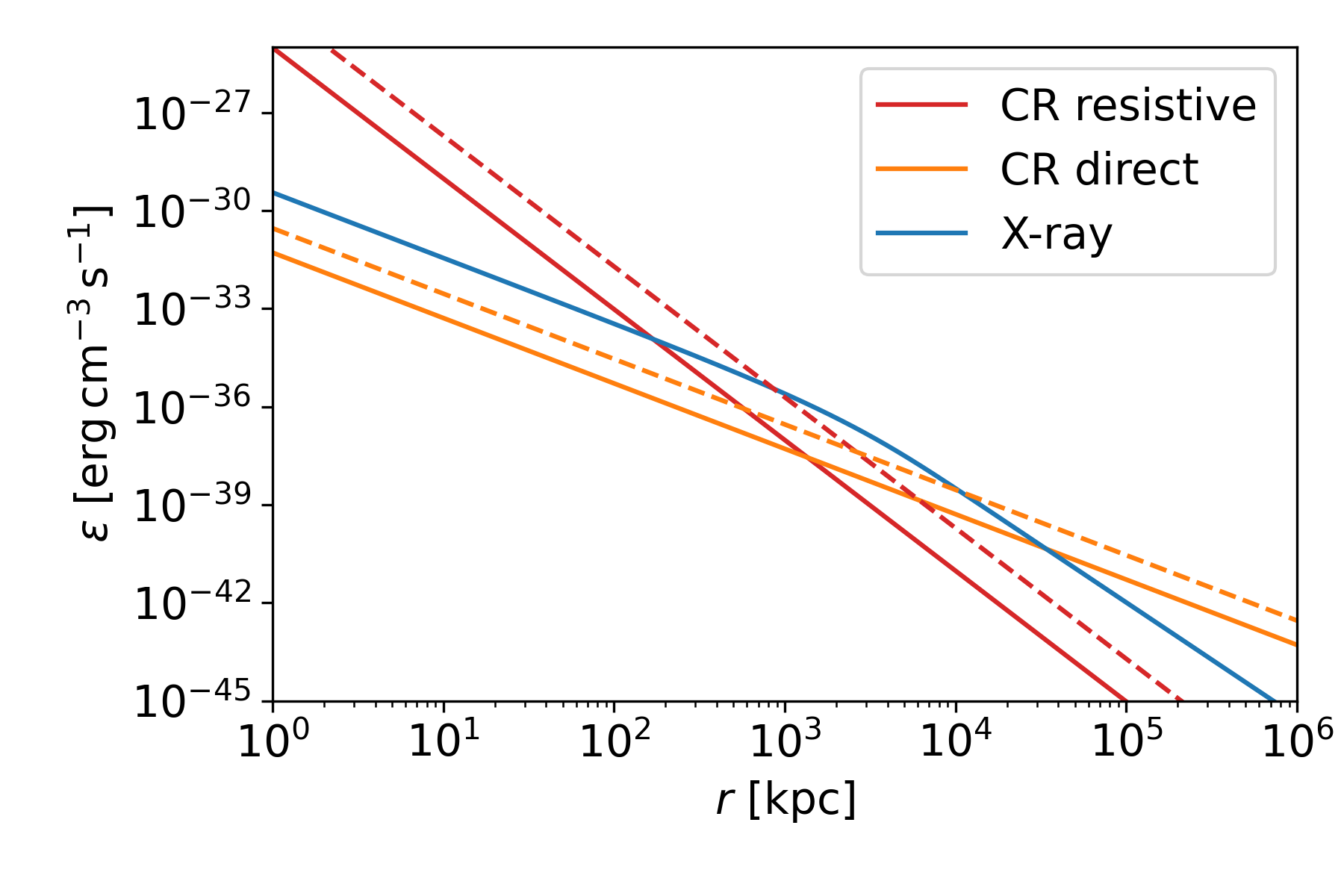}
    \caption{Heating rates per unit volume $\varepsilon$ as a function of distance $r$ from a galaxy.
    The solid red, orange, and blue lines show the heating rates by CR resistive, CR direct, and X-ray heating, respectively, for our fiducial parameters.
    The dashed lines correspond to the heating rates when the  CR spectrum is soft ($s=2.4$ and $E_{\rm max}=0.3 \, \mathrm{GeV}$).
    The parameters used are summarized in Table \ref{tab:parameters}.
    }
    \label{fig:heating}
\end{figure}

We obtained in the previous section \ref{subsec32} the heating rate densities of three mechanisms: X-ray (\ref{eq:Xray}), CR resistive (\ref{eq:CRresistive}), and CR direct heating (\ref{eq:CRdirect}).
Using these equations, we can compare these heating rates as a function of the distance $r$ from a galaxy. 
The result is shown in Figure \ref{fig:heating}, and the parameters used are summarized in Table \ref{tab:parameters}.
The red, orange, and blue solid lines show the heating rate of CR resistive, CR direct, and X-ray heating, respectively.
Figure \ref{fig:heating} clearly shows that in the vicinity of the source, the CR resistive heating is dominant, while in the far outside, CR direct heating has the highest heating rate density.
In between, X-rays can contribute to the heating.
As is also evident in Figure \ref{fig:heating}, the heating rates of CR resistive and direct heating decrease with distance as $\propto r^{-4}$ and $\propto r^{-2}$, respectively.
The radial dependence of X-ray heating changes from $\propto r^{-2}$ to $r^{-3}$ where the optical depth of the lowest energy photon becomes order unity $\tau_{\nu_{\rm min}} \simeq 1$.
The dashed lines in Figure \ref{fig:heating} show the dependence on the spectral index and maximum energy of CRs.
The CR spectral index is changed from $s=2$ to $s=2.4$, as the observations on the Earth imply.
In addition, the maximum energy of CRs is reduced by one order of magnitude ($E_{\rm max}=0.3 \, \mathrm{GeV}$) because maximum CR energy in the high-$z$ galaxies is not fully understood.
This shows that CR heating strongly depends on the spectrum and maximum energy.
The heating rate of the CR resistive heating becomes about $20$ times larger than that in the fiducial case.
This is because, when the CR luminosity is fixed, the CR current density is increased as the number of lower energy CRs increases, while the direct heating also becomes efficient because of the larger cross-sections for lower energy CRs (Equations (\ref{eq:lossCoulomb}), (\ref{eq:lossIon})).

\begin{table}
\caption{Parameters used for Figure \ref{fig:heating}.
}
\label{tab:parameters}
\centering
\begin{tabular}{c|cc}
\hline
Parameters                         & Fiducial & Soft CR spectrum \\ \hline
$T_0 \, (\mathrm{K})$          & $1$                                 & $1$                                  \\
$n_{\rm H} \, (\mathrm{cm^{-3}})$          & $10^{-4}$                                 & $10^{-4}$                                  \\
$\chi_e$                           & $10^{-3}$                                 & $10^{-3}$                                  \\
$f_{\rm X}$                        & $1$                        & $1$                                        \\
$h\nu_{\rm min} \, (\mathrm{keV})$ & $0.2$                                     & $0.2$                                      \\
$h\nu_{\rm max} \, (\mathrm{keV})$ & $10$                                      & $10$                                       \\
$\alpha$                           & $1.5$                                     & $1.5$                              \\
$\varepsilon_{\rm CR}$             & $0.1$                      & $0.1$                                      \\
$E_{\rm min} \, (\mathrm{GeV})$   & $3 \times 10^{-3}$                         & $3 \times 10^{-3}$                          \\
$E_{\rm max} \, (\mathrm{GeV})$   & $3$                                       & $0.3$                                      \\
$s$                                & $2$                                       & $2.4$                                      \\ \hline
\end{tabular}
\end{table}
%

%
%

%
%

\begin{figure*}
    \centering
    \includegraphics[width=160mm]{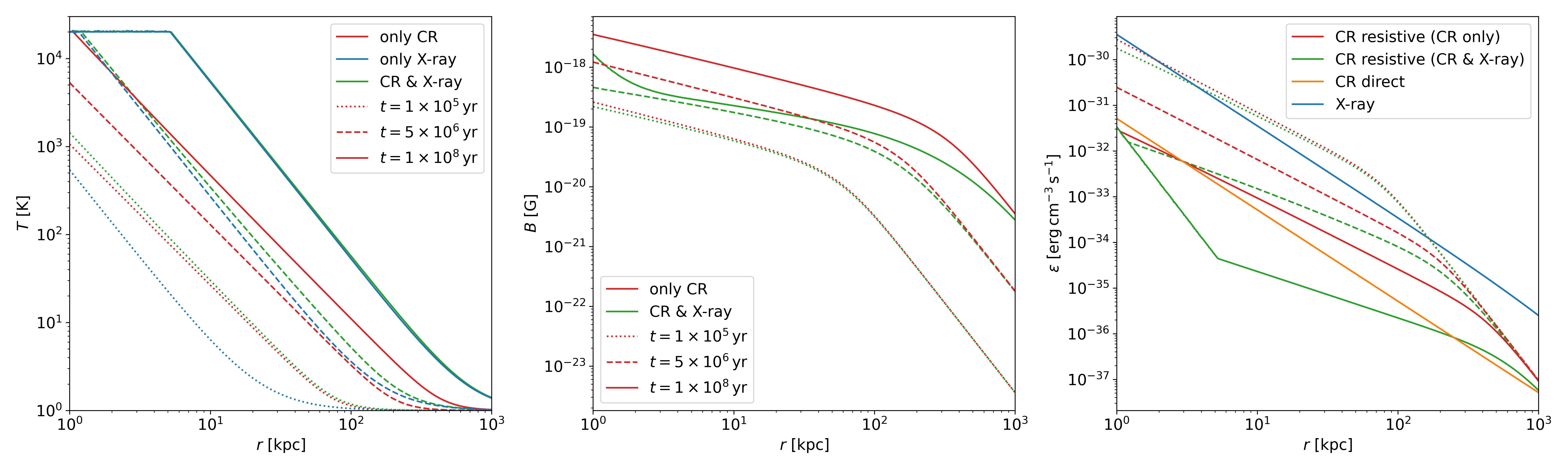}
    \caption{
    The temperature distribution (left), magnetic field strength (middle), and the instantaneous heating rates (right) at $t= 10^5 \, \text{(dotted)}, \, 5 \times 10^6 \, \text{(dashed)}, \, \text{and} \, 10^8 \, \mathrm{yr} \, \text{(solid)}$.
    The red lines in the left panel show the temperature increase only by CR heating, while the blue lines describe the temperature evolution only by X-ray heating.
    The middle panel shows the magnetic field achieved by the mechanism of \citet{Miniati2011}, changing the temperature evolution corresponding to those in the left panel.
    Because the magnetic field generation does not occur in the absence of resistive heating, the magnetic field is not shown when only X-ray heating is included.
    Note that the degree of ionization is fixed to $\chi_{\rm e}$, although X-ray heating depends on $\chi_{\rm e}$.
    }
    \label{fig:evolution}
\end{figure*}

The resistive heating rate depends on the gas temperature through the resistivity (\ref{eq:resistivity}).
We can obtain the temperature increase achieved by CRs and X-rays by integrating the equation
\begin{equation}
    \frac{\mathrm{d}}{\mathrm{d} t} \left( \frac{3}{2} (1 + \chi_{\rm e}) n_{\rm H} k_{\rm B} T \right) = \varepsilon_{\rm X} + \varepsilon_{\rm CR, res} + \varepsilon_{\rm CR, dir}.
\end{equation}
%
Figure \ref{fig:evolution} shows the result of the numerical integration of this.
We set the initial temperature to be $T_0 = 1 \, \mathrm{K}$ everywhere, and the other parameters including density $n_{\rm H}$, ionization fraction $\chi_{\rm e}$ are fixed to the fiducial ones in the Table \ref{tab:parameters}.
The temperature increase is stopped at $T=2 \times 10^4 \, \mathrm{K}$ by hand, assuming that above this temperature, H$_{\rm I}$ atomic cooling works efficiently.
The left panel shows the temperature at $t=10^5 \, \text{(dotted)}, \, 
5 \times 10^6 \, \text{(dashed)}, \, \text{and} \, 10^8 \, \mathrm{yr} \, \text{(solid)}$.
The red lines in the left panel show the temperature when only CRs contribute to the temperature increase, while the blue lines show the temperature when only X-rays raise the temperature.
The green lines correspond to the case where both CRs and X-rays contribute to the temperature increase.
The right panel shows the instantaneous heating rates.
While the heating rates of CR direct and X-ray heating (orange and blue lines) do not change, the CR resistive heating rate changes in time because it depends on the temperature.
The red and green lines correspond to the heating rates at $t=10^5 \, \text{(dotted)}, \, 
5 \times 10^6 \, \text{(dashed)}, \, \text{and} \, 10^8 \, \mathrm{yr} \, \text{(solid)}$ in the case only CRs raise the temperature, and in the case both X-rays and CRs increase the temperature, respectively.
The middle panel shows the magnetic field strength generated by the mechanism by \citet{Miniati2011}, and we will discuss this in the next subsection \ref{subsec34}.

The abrupt temperature increase at earlier times is attributed to the high heating rate of CR resistive heating.
However, the heating rate decreases fast as the temperature increases.
After all, the temperature increase up to $T \sim 10^4 \, \mathrm{K}$ is achieved mainly by X-rays.
The temperature reaches $10^4 \, \mathrm{K}$ at $r = 1\, \mathrm{kpc}$ at $t\simeq 2 \times 10^6 \, \mathrm{yr}$.
Note that heating rates by X-ray and CR direct heating are not changed because we fixed the density, ionization degree, and fluxes of CRs and X-rays.
This point will be discussed in section \ref{sec:4}.

\begin{figure*}
    \centering
    \includegraphics[width=160mm]{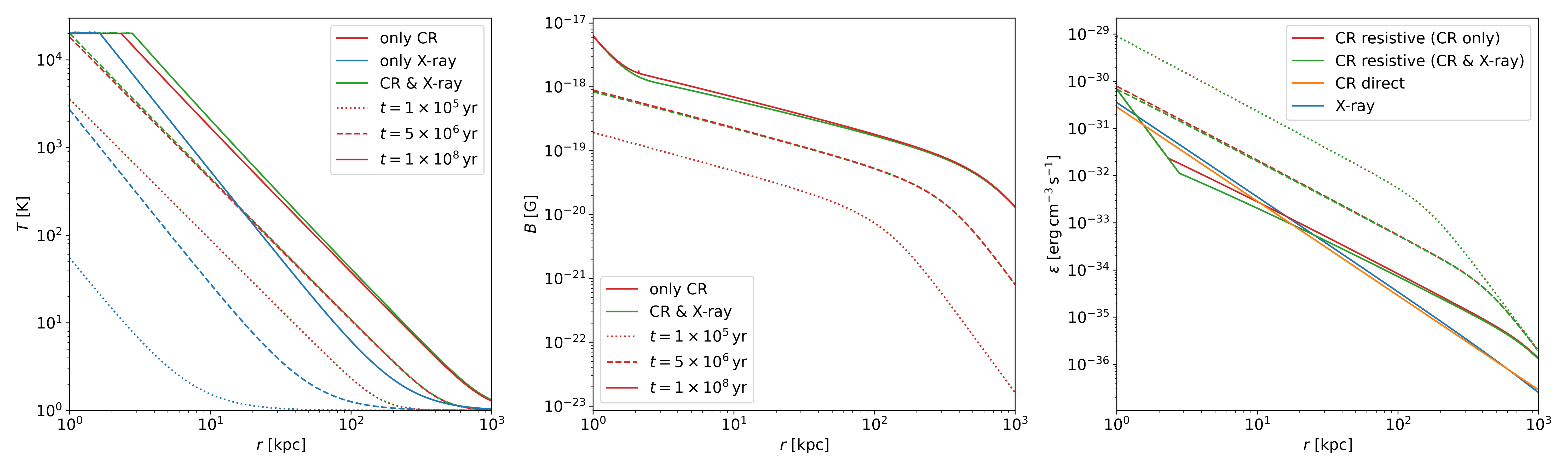}
    \caption{
    The same as Figure \ref{fig:evolution}, but the 
    spectral index and maximum energy of CRs are changed to $s=2.4$ and $E_{\rm max} = 0.3 \, \mathrm{GeV}$.
    In addition, X-ray luminosity is reduced to $f_{\rm X} = 0.1$
    }
    \label{fig:CRdominant}
\end{figure*}

We also plotted in Figure \ref{fig:CRdominant} the temperature evolution for the case when the CR spectrum is soft and X-ray luminosity is low, that is, parameters are changed from fiducial ones to $s=2.4, \, E_{\rm max}=0.3 \, \mathrm{GeV}$, and $f_{\rm X}=0.1$.
This clearly shows that CR resistive heating is still dominant at $t=10^8 \, \mathrm{yr}$.
Because the X-ray luminosities and CR spectrum are highly uncertain in high-$z$ galaxies, such a situation might be realized in the early Universe.

\subsection{Magnetic field generation}
\label{subsec34}
It is pointed out that streaming CRs can generate magnetic fields around a galaxy \citep{Miniati2011, Ohira2020, Ohira2021, Yokoyama2022}.
As investigated in \cite{Yokoyama2022}, for the length scale greater than $\sim 1 \mathrm{kpc}$ and initial temperature $T_0 \lesssim 10 \, \mathrm{K}$, the magnetic field generation proposed by \cite{Miniati2011} is the dominant mechanism.
However, in \cite{Miniati2011} and \cite{Yokoyama2022}, only CR resistive heating is considered as a heating process.
Therefore, we recalculated the magnetic field generation, including the CR resistive, CR direct, and X-ray heating.
The magnetic field generated by the mechanism of \citet{Miniati2011} is shown in the middle panels of Figure \ref{fig:evolution} and \ref{fig:CRdominant}.
The equation used to calculate the magnetic field is:
\begin{equation}
    \frac{\mathrm{d} B}{\mathrm{d} t} = \frac{c \eta_{\rm C}}{L} J_{\rm CR},
    \label{eq:MBcal}
\end{equation}
where $\eta_{\rm C}$ is the Spitzer resistivity (\ref{eq:resistivity}) and a function of the temperature $T$ \citep[see][for the detail]{Yokoyama2022}.
$L$ is the length scale of the temperature gradient, and we set $L=r$, for simplicity.
The red lines in the middle panels of Figure \ref{fig:evolution} and \ref{fig:CRdominant} show the magnetic field strength when only CRs heating is included, 
while the green lines are the strength when both CR and X-ray heating are included.
Because the magnetic field generation does not work in the absence of resistive heating, the magnetic field evolution is not shown when only X-ray heating is included.
These figures tell us that the growth of magnetic fields can be suppressed by X-ray heating by about $1$ order of magnitude.
This is because the resistivity and resistive electric field are large when the temperature is low.
Due to the quick temperature increase by X-ray heating, the resistive electric field rapidly gets weak, making the generated magnetic field weaker.
Note that we did not plot the magnetic fields generated by the mechanism of \cite{Ohira2021} and \cite{Yokoyama2022} because they are weaker than that generated by the mechanism by \cite{Miniati2011} for these parameters.


\section{Discussion}
\label{sec:4}
As mentioned in \ref{subsec33}, we kept all the parameters other than the gas temperature constant.
Because CR direct heating and X-ray heating include the ionization process, the degree of ionization must be changed as time goes on. 
In addition, collisional ionization by thermal particles may come in as the temperature raises.
Because the X-ray heating depends on the ionization degree, X-ray heating may be suppressed as the temperature increases.
We also neglected energy loss processes and scattering of CRs during their propagation.
Due to the energy loss, CRs get unable to propagate far from the galaxy and the CR current density becomes small faster than $\propto r^{-2}$.
Therefore, self-consistent treatment of heating, ionization, gas dynamics, and CR and photon propagation is necessary to determine the actual temperature increase and this will be investigated in the future study.
Another important process to consider during CR propagation is kinetic plasma instabilities.
It is expected that the system with streaming CRs is unstable.
Because of the plasma instabilities, CR energy can be transferred to the gas via electromagnetic fields.
In order to understand this heating process, it is necessary to know the nonlinear evolution of plasma instabilities.
Therefore, we have to conduct plasma particle simulations.

In this work, we assumed that thermal electrons are responsible for the return current.
However, if we consider the secondary electrons ejected by the ionization of neutral hydrogen by CRs, these secondary electrons might be subject to runaway acceleration by the resistive electric field \citep[][]{Ohira2022}.
In this case, the return current is replaced by the accelerated secondary electrons, and because of the high energy of the secondary electrons, the Coulomb collision becomes less efficient, reducing the resistive electric field and heating.
Whether or not the runaway acceleration occurs and how large resistive electric fields remain after that depend on the situation.
In addition, X-rays also produce secondary electrons by photoionization.
The self-discharge in the realistic system should be investigated further.

Another important effect caused by the streaming CRs is magnetic field generation \citep[][]{Miniati2011, Ohira2020, Ohira2021, Yokoyama2022}.
\citet{Bell2003, Miniati2011} discussed the magnetic field generation by inhomogeneous resistive heating induced by CRs, which is described by the following equation
\begin{equation}
    \pd{\bm{B}}{t} = c\nabla \times (\eta_{\rm C} \bm{J}_{\rm CR}).
    \label{eq:MB}
\end{equation}
On the other hand, \citet{Ohira2021} and \citet{Yokoyama2022} discussed the Biermann battery which is triggered by the pressure gradient produced by streaming CRs in two different mechanisms.
The magnetic field generation by the Biermann battery is described by \citep[][]{Biermann1950}
\begin{equation}
    \pd{\bm{B}}{t} = -\frac{c}{en_{\rm e}^2} \nabla n_{\rm e} \times \nabla P_{\rm e},
    \label{eq:Biermann}
\end{equation}
where $P_{\rm e}$ is the electron pressure.
The electron pressure gradient is induced by inhomogeneous resistive heating in the case of \citet{Yokoyama2022} and by inhomogeneous adiabatic compression of electron fluid in the case of \citet{Ohira2021}.
This pressure gradient is not necessarily parallel to the density gradient of the ambient medium, thereby generating nonzero $\nabla n_{\rm e} \times \nabla P_{\rm e}$ and magnetic field.
The dominance of the above three mechanisms depends on length scale, temperature, density, ionization degree, and so on \citep[][]{Yokoyama2022}.
In addition, the generated magnetic field might affect the propagation of CRs, and this influences the heating and ionization rates by CRs.
Therefore, the propagation of CRs and X-rays, the time evolution of the temperature, density, and ionization degree of gas, and the generation of magnetic fields should be solved self-consistently.

Although we have focused on the heating outside the galaxy in this work, CR heating may also work inside the galaxy \citep[e.g.][]{Owen2019}.
Because CR resistive heating works efficiently in low-temperature regions, it may change the temperature of gas clouds and filaments and affect the star formation processes.
In addition, the ionization degree of dense clouds might be determined by the ionization by CRs.
Because the ionization degree determines the coupling between the gas and magnetic fields, this can also influence the formation of stars \citep[e.g.][]{McKee2007}.
If the magnetic field in the galaxy is sufficiently strong, CRs exert pressure on the gas.
It is known that CR pressure is responsible for the driving of galactic winds and this might pollute the pristine gas with metals \citep[e.g.][]{Shimoda2022}.
Although metal enrichment in the early Universe is vigorously investigated \citep[e.g.][]{Chiaki2018}, CRs might also contribute to the transport of metals and this will affect the heating and cooling of the gas.
CRs can produce \citep{Ohira2020, Ohira2021, Yokoyama2022} and amplify magnetic fields \citep[e.g.][]{Bell2004} and, in turn, magnetic fields affect the propagation of CRs.
Therefore, CRs are an essential component in understanding the galaxy evolution in the early Universe.

\section{Summary}
\label{sec:5}
We revisited the influence of streaming CRs on the heating of the IGM.
We clarified that resistive heating induced by CRs becomes dominant over the other mechanisms in the vicinity of a galaxy at $r\lesssim 10^2 \, \mathrm{kpc}$ in early phases.
The temperature reaches $10^4 \, \mathrm{K}$ at $r=1 \, \mathrm{kpc}$ in $2 \times 10^{6} \, \mathrm{yr}$.
Because the CR heating depends on its amount, minimum and maximum energy, and the spectrum of the accelerated CRs (see Figure \ref{fig:heating}), it is important to understand the CR acceleration in the early Universe.
CRs affect the evolution of the gas through heating, magnetic field generation and amplification, and metal transport, thereby affecting the subsequent star formation.
Therefore, it is important to reveal the role of CRs in order to understand the galaxy evolution in the early Universe correctly.
The signature of IGM heating by CRs may be found by future $21$-$\mathrm{cm}$ line observations because CR heating has a characteristic length scale different from that of X-ray heating.
The correct understanding of CR heating will lead to the verification of the presence of the first CRs.

\section*{acknowledgments}
We thank M. Hoshino and G. Chiaki for the useful discussions.
This work is supported by JSPS KAKENHI grant numbers JP21J20737 (SY), JP19H01893 (YO), and JP21H04487 (YO).

\section*{Data availability}
No new data were generated or analysed in support of this research.

\bibliographystyle{mnras}
\bibliography{library}

\bsp	
\label{lastpage}
\end{document}